\documentclass[aps,pra,superscriptaddress,twocolumn]{revtex4-1}

\usepackage[utf8,latin1]{inputenc}
\usepackage[T1]{fontenc}     
\usepackage[american,ngerman,greek,brazilian,british]{babel}
\usepackage{float}
\usepackage[dvipsnames]{xcolor}
\usepackage[colorlinks=true,citecolor=Aquamarine,linkcolor=Mulberry,urlcolor=Magenta,hyperindex]{hyperref}
\usepackage{cleveref}
\usepackage{graphicx}
\usepackage{epsfig}
\usepackage{color}
\usepackage{lmodern}
\usepackage{mathpazo}
\usepackage[babel=true]{microtype}
\usepackage{amsmath,amssymb,amsthm}
\usepackage{mathtools}
\usepackage{bm}
\usepackage{enumerate}
\usepackage{sidecap}
\usepackage{enumitem}
\usepackage{listings}

\newcommand{\ket}[1]{\vert#1\rangle}

\newcommand{\braket}[2]{\langle#1\vert#2\rangle}
\newcommand{\ketbra}[2]{\vert #1 \rangle \langle #2 \vert}

\newcommand{\1}{\mathbb{1}}
\newcommand{\set}[1]{\mathcal{#1}}

\newcommand{\code}[1]{\texttt{#1}}

\newcommand{\tr}{\text{\normalfont Tr}}

\newcommand{\beq}{\begin{equation}}
\newcommand{\eeq}{\end{equation}}

\newcommand{\ufmgfis}{Departamento de F\'isica, Universidade Federal de Minas Gerais, Caixa Postal 702, 31270-901, Belo Horizonte, MG, Brazil}
\newcommand{\ufmgmat}{Departamento de Matem\'atica, Universidade Federal de Minas Gerais, Caixa Postal 702, 31270-901, Belo Horizonte, MG, Brazil}
\newcommand{\icfo}{ICFO-Institut de Ciencies Fotoniques, The Barcelona Institute of Science and Technology, 08860 Castelldefels (Barcelona), Spain}

\newcommand{\unicamp}{Departamento de Matem\'atica Aplicada, IMECC-Unicamp, 13084-970, Campinas, S\~ao Paulo, Brazil}
\newcommand{\todai}{Department of Physics, Graduate School of Science, The University of Tokyo, Hongo 7-3-1, Bunkyo-ku, Tokyo 113-0033, Japan}
\newcommand{\iqoqi}{Institute for Quantum Optics and Quantum Information (IQOQI), Austrian Academy of Sciences, Boltzmanngasse 3, A-1090 Vienna, Austria}

\begin{document}

\title{Most incompatible measurements for robust steering tests} 

\author{Jessica Bavaresco}
\email{jessica.bavaresco@oeaw.ac.at}
\affiliation{\ufmgfis}
\affiliation{\iqoqi}

\author{Marco T\'ulio Quintino}
\affiliation{\todai}

\author{Leonardo Guerini}
\affiliation{\ufmgmat}
\affiliation{\icfo}

\author{Thiago O. Maciel}
\affiliation{\ufmgfis}

\author{Daniel Cavalcanti}
\affiliation{\icfo}

\author{Marcelo Terra Cunha}
\affiliation{\unicamp}

\date{\today}

\begin{abstract}
We address the problem of characterizing the steerability of quantum states under restrictive measurement scenarios, i.e., the problem of determining whether a quantum state can demonstrate steering when subjected to $N$ measurements of $k$ outcomes. We consider the cases of either general positive operator-valued measures (POVMs) or specific kinds of measurements (e.g., projective or symmetric). We propose general methods to calculate lower and upper bounds for the white-noise robustness of a $d$-dimensional quantum state under different measurement scenarios that are also applicable to the study of the noise robustness of the incompatibility of sets of unknown qudit measurements. 
We show that some mutually unbiased bases, symmetric informationally complete measurements, and other symmetric choices of measurements are not optimal for steering isotropic states and provide candidates to the most incompatible sets of measurements in each case. 
Finally, we provide numerical evidence that nonprojective POVMs do not improve over projective ones for this task.
\end{abstract}

\maketitle

\section{Introduction}\label{intro}

Correlations arising from local measurements on entangled states can lead to statistics that cannot be explained by any local causal theory \cite{RevNL}. This nonlocal aspect of quantum mechanics can be analyzed from two nonequivalent perspectives, Bell nonlocality \cite{Bell} and Einstein-Podolsky-Rosen (EPR) steering \cite{EPR,Wiseman}. On a bipartite scenario, while Bell nonlocality deals with a full device-independent approach where a correlation experiment is analyzed only considering the probability relations between inputs and outcomes, EPR steering plays an intermediate role between entanglement and Bell nonlocality by doing a device-independent analysis only on one side of the experiment while treating the other side in a device-dependent manner (e.g., performing full state tomography).

Although the first notions of EPR steering date back to 1935 \cite{EPR}, its modern mathematical formulation only appeared in 2007 \cite{Wiseman}, and many of its fundamental properties \cite{MTGeneralMeas,MTOneWay,MultipartiteDani,aolita14,quintino17} and applications to semi-device independent protocols \cite{tomamichel11,tomamichel12,OneSidedDIQC,law14,passaro15,kogias17} are only being understood recently. In order to get a better understanding of EPR steering and make use of its practical applications, an important task is to determine which states can lead to these nonlocal correlations. 

EPR steering can be certified with the use of steering witnesses \cite{SteIneq}, but finding suitable inequalities and choosing appropriate measurements to reveal this sort of nonlocality of a given quantum state remains a nontrivial task. On the other hand, proving that a quantum state cannot lead to EPR steering can be done by constructing a local hidden state (LHS) model that is able to simulate the statistics of the quantum state \cite{WernerModel,Wiseman,sania14,JoeCriterion}. Recently, a general algorithm to construct LHS models for quantum states was presented \cite{LHScavalcanti,LHShirsch}, regarding scenarios where all (i.e., infinitely many) measurements are considered. However, from a practical perspective, it is important to characterize what one can do with a limited number of measurements and outcomes, or yet, when even the structure of the allowed measurements is restricted.

Since the ability to demonstrate steering is intimately related to the ability to perform incompatible measurements \cite{krausbook,buschbook}, by addressing the problem of characterizing the steerability of quantum states under restrictive scenarios one can simultaneously address the problem of characterizing the ability to jointly perform a set of unknown measurements subjected to the same restrictions \cite{MTJM,UolaJM1}. Although the question of whether a set of fixed (known) measurements is jointly measurable can be decided by semidefinite programming (SDP) \cite{wolf09} -- which is also the case for deciding whether a given quantum state is steerable when subjected to a set of fixed (known) measurements \cite{MattPusey,SteeringWeight} -- if the complete description of the measurements is not known, there do not exist general methods to characterize steerability or joint measurability.

In this paper we consider steering and joint measurability in scenarios where the number of measurements and outcomes is finite. By systematically applying adaptations of the parametric search, the seesaw algorithm \cite{RevSDP}, and the outer polytope approximation \cite{SuperPOVM}, we derive upper and lower bounds to the maximal amount of white noise that a quantum state can endure before it is no longer able to demonstrate steering when subjected to a set of $N$ general local measurements with $k$ outcomes. Using the same methods, we calculate upper and lower bounds for the minimum amount of white noise that must be applied to \textit{any} set of general $N$ qudit $k$-outcome measurements so that it is assured that they can be jointly performed. By imposing further restrictions on the measurement scenarios, we also study prominent classes of measurements that are known to be useful in many quantum information tasks, such as projective measurements, symmetric informationally complete (SIC) measurements \cite{renes04}, and measurements from mutually unbiased bases (MUBs) \cite{durt10}. 

We present our calculations for qubit states in scenarios ranging from $2$ to $18$ projective, planar projective, symmetric, and general measurements and provide strong evidence that, in the considered scenarios, general positive-operator valued measures (POVMs) do not outperform projective measurements for exhibiting steering. We also show that our optimal sets of qubit measurements are not distributed in the Bloch sphere according to any of the most intuitive patterns, such as the vertices of Platonic solids, the distribution of electrons on a sphere in the Thomson problem \cite{Thomson}, and the Fibonacci spiral \cite{Fibonacci}. We present an alternative candidate for this distribution for the cases of $N\in\{2,\ldots,6\}$ measurements, supported by our numerical findings. For higher dimensions, we present evidence that increasing the number of outcomes beyond the value of the local dimension of the state does not improve white-noise robustness, again implying that projective measurements are optimal for steering. We also prove that, in many cases, incomplete sets of MUB measurements are not optimal, while providing numerical evidence that complete sets may be optimal for steering isotropic states.

\section{Preliminaries}

\subsection{Einstein-Podolsky-Rosen steering}

Bipartite steerability is usually defined in terms of an assemblage. Let $\rho_{AB}$ be a bipartite quantum state shared by Alice and Bob and let $\{M_{a|x}\}$ be a set of measurements on Alice's subsystems. Then, an assemblage $\{\sigma_{a|x}\}$ is defined as
\beq
\sigma_{a|x} = \tr_A(M_{a|x}\otimes\1\,\rho_{AB}),
\eeq
for all $a,x$, where $x\in\{1,\ldots,N\}$ and $a\in\{1,\ldots,k\}$ label Alice's measurements and outcomes, respectively, and $\tr_A$ denotes the partial trace over the Hilbert space of Alice. An assemblage does not demonstrate steering when it admits an LHS model, namely, when there exists $\Lambda$ such that
\beq\label{eqlhs}
\sigma_{a|x} = \sum_\lambda \pi(\lambda)p_A(a|x,\lambda)\rho_\lambda,
\eeq
for all $a,x$, where $\lambda\in\Lambda$ are the possible values that can be assumed by a local hidden variable with probability $\pi(\lambda)$, $p_A(a|x,\lambda)$ is the probability of Alice's obtaining outcome $a$ conditioned on her choice of measurement $x$ and $\lambda$, and, finally, $\rho_\lambda$ is a local hidden state held by Bob that is conditioned by the value $\lambda$ and independent of Alice's measurements and outcomes. An assemblage demonstrates steering when it does not admit such decomposition \cite{Wiseman} or, equivalently, when it violates a steering inequality \cite{SteIneq}. A quantum state $\rho_{AB}$ is unsteerable if all assemblages that can be generated by performing local measurements on it admit an LHS model. On the other hand, a quantum state is steerable if there exists a set of measurements that, when locally performed on it, generates an assemblage that violates a steering inequality.

 \subsection{Measurement incompatibility}  
 
A set of measurements $\{M_{a|x}\}$, where $x\in\{1,\ldots,N\}$ labels the measurements in the set and $a\in\{1,\ldots,k\}$ labels the outcomes of each measurement, is jointly measurable, or compatible, if there exists a \textit{joint measurement}, $\{M_\lambda\}$, such that
\begin{equation}
M_{a|x} = \sum_\lambda \pi(\lambda) p(a|x,\lambda) M_\lambda,
\end{equation}
for all $a,x$, where $\pi(\lambda)$ and $p(a|x,\lambda)$ are elements of probability distributions. Hence, all POVM elements $M_{a|x}$ can be recovered by coarse-graining over the joint measurement $\{M_\lambda\}$. 

Although for projective measurements joint measurability is equivalent to commutation, general POVMs from a jointly measurable set may not commute \cite{kru,teiko08}. In this sense, joint measurability is a more general definition of incompatibility.

 \subsection{Main problem} \label{Quantify}

Consider the depolarizing map $\Lambda_\eta$ acting on the Hermitian operator $A$ of a $d$-dimensional Hilbert space $\set{H}$, defined as
\beq\label{eqdepolmap}
A\mapsto\Lambda_{\eta}(A) = \eta A+(1-\eta)\tr(A)\frac{\1}{d}.
\eeq
The depolarizing map can be physically interpreted as the effect of the presence of white noise in the implementation of $A$.
When applied to elements of an assemblage it defines a steering quantifier, the white-noise robustness of an assemblage,
\beq
\eta(\{\sigma_{a|x}\})=\max\left\{\eta \ | \ \{\Lambda_{\eta}(\sigma_{a|x})\}_{a,x} \in LHS\right\},
\eeq 
where $LHS$ is the set of assemblages that admit an LHS model and, hence, do not demonstrate steering. Therefore, $\eta(\{\sigma_{a|x}\})$ is the exact value of $\eta$, called the critical visibility of the assemblage, above which $\{\sigma_{a|x}\}$ no longer admits an LHS model. Since $\{\Lambda_1(\sigma_{a|x})\}$ is the assemblage itself and $\{\Lambda_0(\sigma_{a|x})\}$ is such that each of its elements corresponds to a multiple of the identity (and therefore it always admits an LHS model), by convexity it is guaranteed that the critical visibility of the assemblage $\eta(\{\sigma_{a|x}\})$ lies in $[0,1]$.

Given an assemblage, its critical visibility can be calculated by an SDP (see Sec. \ref{upper} and see Ref. \cite{RevSDP} for a review of SDP characterization of steering). Similarly, by applying the depolarizing map to a set of measurements $\{M_{a|x}\}$ instead of an assemblage, one can define the critical visibility for a set of measurements to be incompatible, i.e., a value of $\eta$ above which a set of measurements can no longer be described by a joint POVM.

Here we are interested in calculating the minimum of the quantity $\eta(\{\sigma_{a|x}\})$ among all the possible choices of $N$ measurements with $k$ outputs for a fixed quantum state $\rho_{AB}$. Formally this quantity can be defined as
\beq
\begin{split}
\eta^*(\rho_{AB},N,k) = \min_{\{M_{a|x}\}} \Big\{ &\eta(\{\sigma_{a|x}\}) \ | \\
 								 &\sigma_{a|x}=\tr_A(M_{a|x}\otimes\1\,\rho_{AB}) \Big\},
\end{split}
\eeq
where the minimization runs over sets $\{M_{a|x}\}$ of $N$ $k$-outcome measurements. The value $\eta^*(\rho_{AB}, N, k)$ is the critical visibility of the quantum state $\rho_{AB}$ when subjected to $N$ measurements of $k$ outcomes. Note that for $\eta\leq\eta^*(\rho_{AB}, N, k)$, the state $\rho_{AB}$ is unsteerable for {\it all} sets of $N$ $k$-outcome measurements, and for $\eta>\eta^*(\rho_{AB}, N, k)$, $\rho_{AB}$ is steerable for {\it at least one} set of $N$ $k$-outcome measurements. 

Unlike the critical visibility of an assemblage $\eta(\{\sigma_{a|x}\})$, the critical visibility of a quantum state $\eta^*(\rho_{AB}, N, k)$ is the solution of a min-max optimization problem and cannot be calculated by an SDP. In this work we provide methods to obtain upper and lower bounds for $\eta^*(\rho_{AB},N,k)$.

\subsection{Connection to the most incompatible measurements}\label{JM=STE}

In Refs. \cite{MTJM,UolaJM1} the authors have proved that a set of measurements $\{M_{a|x}\}$ is not jointly measurable \textit{if and only if} Alice can steer Bob by performing the same measurements on her share of a maximally entangled state $\ket{\phi_d^+}:=\frac{1}{\sqrt{d}}\sum_{i=0}^{d-1}\ket{ii}$, where $d$ stands for the local dimension of the quantum system. Hence, the critical visibility $\eta^*(\ket{\phi_d^+}, N,k)$ coincides with the critical visibility for which \textit{any} set of $N$ measurements with $k$ outcomes is jointly measurable. 

Following from the definition of the depolarizing map (Eq. (\ref{eqdepolmap})) and the maximally entangled states, it is easy to show that the noisy assemblage $\{\Lambda_\eta(\sigma_{a|x})\}$, resulting from applying the depolarizing map to an assemblage generated by performing local measurements $\{M_{a|x}\}$ on a maximally entangled state $\ket{\phi_d^+}$, is equivalent to the assemblage resulting from locally performing measurements $\{M_{a|x}\}$ on the noisy state $(\1\otimes\Lambda_\eta)(\ketbra{\phi_d^+}{\phi_d^+})$. Namely, 
\beq
\Lambda_\eta(\sigma_{a|x})=\tr_A(M_{a|x}\otimes\1_d\,(\1_d\otimes\Lambda_\eta)(\ketbra{\phi_d^+}{\phi_d^+})),
\eeq
where 
\beq
(\1_d\otimes\Lambda_\eta)(\ketbra{\phi_d^+}{\phi_d^+})=\eta\ketbra{\phi_d^+}{\phi_d^+}+(1-\eta)\frac{\1_{d^2}}{d^2}
\eeq
is the isotropic state of local dimension $d$. Therefore, the critical visibility of the maximally entangled states $\eta^*(\ket{\phi_d^+},N,k)$ is equal to the critical value of the parameter $\eta$ of the isotropic states for which they can demonstrate steering, which, in turn, is equal to the critical visibility for any set of $N$ unknown qudit measurements with $k$ outcomes to be compatible. For this reason we speak equivalently of the critical visibility of the maximally entangled states, isotropic states, and joint measurability. To simplify notation, we define this quantity as $\eta^*(d, N,k)\coloneqq\eta^*(\ket{\phi_d^+}, N,k)$. 

For general states, one can lower-bound the noise robustness of joint measurability by that of steerability \cite{QuantifiersDani}.

\section{Methods}\label{methods}

In the following we describe three methods we used to characterize the steerability of quantum states subjected to restricted measurement scenarios. The first method provides upper bounds for $\eta^*(\rho_{AB}, N,k)$ in scenarios where not only the number of measurements and outcomes is fixed but possibly also the structure of the POVMs. The second one provides upper bounds for $\eta^*(\rho_{AB},N,k)$ when only the number of measurements and outcomes is fixed (considering general measurements). Both methods provide candidates for the optimal set of measurements in a given scenario. The third method provides lower bounds for $\eta^*(\rho_{AB},N,k)$ and constructs LHS models for quantum states when the number of measurements and outcomes is fixed.

All code used in this work is available in Ref. \cite{Code}.

\subsection{Upper bounds for $\eta^*(\rho_{AB},N,k)$}\label{upper}
\textit{Search algorithm}. For a given quantum state $\rho_{AB}$ and a fixed set of measurements $\{M_{a|x}\}$, the critical visibility $\eta(\{\sigma_{a|x}\})$ of the assemblage $\{\sigma_{a|x}\}$, which is generated by locally performing these measurements on the given state, can be calculated by SDP
	\begin{align}
	\begin{split}\label{wnrsdp}
	\text{given}                                   &\hspace{0.2cm} \rho_{AB}, \{M_{a|x}\} \\
	\max	  					  &\hspace{0.2cm} \eta  \\
	\text{s.t.}					  &\hspace{0.2cm} \sigma_{a|x} = \tr_A(M_{a|x}\otimes\1 \, \rho_{AB}),\ \forall\, a,x \\
		 					  & \hspace{0.2cm} \eta\sigma_{a|x}+(1-\eta)\tr(\sigma_{a|x})\frac{\1}{d}=\sum_\lambda D(a|x,\lambda)\sigma_\lambda,\ \forall\, a,x \\
							  & \hspace{0.2cm} \sigma_\lambda \geq 0, \ \forall\,\lambda,
	\end{split}
	\end{align}
where $D(a|x,\lambda)$ are elements of deterministic probability distributions and $\lambda\in\{1,\ldots,k^N\}$.
For a fixed quantum state $\rho_{AB}$, different sets of measurements can be tested, each set requiring one SDP to calculate the value of $\eta(\{\sigma_{a|x}\})$. The first method we propose is to parametrize the sets of measurements allowed in a given scenario and, by varying these parameters, explore the solution of multiple SDPs to calculate a bound for $\eta^*(\rho_{AB},N,k)$.

Two important facts can be explored to facilitate this task. The first one is that it is only necessary to optimize over extremal measurements. This is due to the fact that the critical value of $\eta$ depends linearly on the choice of measurements, hence, by convexity, the optimal value will be obtained over extremal measurements. The second fact is that for a system of dimension $d$, extremal measurements have at most $d^2$ outcomes \cite{ExtremalPOVM}.

Aside from the restriction on the number of measurements and outcomes, it is possible to impose restrictions on the parametrization that specify a certain kind of measurement that can be more relevant to the problem one wishes to approach. For instance, it is possible to perform an optimization over only projective measurements or other POVMs with some specific structure (e.g., SIC-POVMs).

The optimization tools chosen for this work are the MATLAB functions \code{fminsearch} \cite{fminsearch}, an unconstrained nonlinear multivariable optimization tool, and \code{fmincon} \cite{fmincon}, a constrained nonlinear multivariable optimization tool. These methods are heuristic and, as such, are not guaranteed to find a global minimum. In order to improve the bound they provide for $\eta^*(\rho_{AB},N,k)$, multiple different initial points can be tested. They also provide a candidate for the optimal set of $N$ $k$-outcome measurements in the given scenario, the one that generates the most robust assemblage when locally performed on $\rho_{AB}$.
\\

\textit{See-saw algorithm}. The seesaw algorithm is an iterative method for solving some nonlinear optimization problems that has found many applications in quantum information theory. In Refs. \cite{CounterPeres,DisprovePeres,RevSDP} seesaw algorithms are used as methods of measurement optimization that are here adapted to approach our problem. 

Our seesaw iterates two SDPs. The first one is the dual formulation of SDP (\ref{wnrsdp}):
	\begin{subequations}
	\begin{align}
	\begin{split}\label{seesaw1}
	\text{given}                                   &\hspace{0.2cm} \rho_{AB}, \{M_{a|x}\} \\
	\min_{\{F_{a|x}\}}			  &\hspace{0.2cm} 1 - \sum_{a,x} \tr(F_{a|x}\sigma_{a|x})  \\
	\text{s.t.}					  &\hspace{0.2cm} \sigma_{a|x} = \tr_A(M_{a|x}\otimes\1 \, \rho_{AB}) ,\ \forall\, a,x \\
		 					  & \hspace{0.2cm} 1 - \sum_{a,x} \tr (F_{a|x}\sigma_{a|x}) + \frac{1}{d} \sum_{a,x} \tr (F_{a|x})\tr(\sigma_{a|x}) = 0 \\
							  & \hspace{0.2cm} \sum_{a,x} D_\lambda(a|x)F_{a|x} \leq 0, \ \forall\,\lambda. 
	\end{split}
	\end{align}
This SDP returns the coefficients $\{F_{a|x}\}$ of a steering inequality of the form $\sum_{a,x}\tr(F_{a|x}\sigma_{a|x})\geq0$. The value obtained by the assemblage $\{\sigma_{a|x}\}$, which is generated by the input state $\rho_{AB}$ and set of measurements $\{M_{a|x}\}$, for the left hand side of this inequality is precisely $1-\eta(\{\sigma_{a|x}\})$. This is due to the fact that primal and dual problems satisfy strong duality. As part of the seesaw, this SDP starts by taking a randomly chosen set of $N$ $k$-outcome measurements and the quantum state whose steerability one wishes to characterize.
	
The second SDP of the seesaw is
	\begin{align}
	\begin{split}\label{seesaw2}
	\text{given}                 			&\hspace{0.2cm} \rho_{AB}, \{F_{a|x}\}  \\
	\max_{\{M_{a|x}\}} 				&\hspace{0.2cm} \sum_{a,x} \tr(F_{a|x}\sigma_{a|x})  \\ 
	\text{s.t.}  						&\hspace{0.2cm} \sigma_{a|x} = \tr_A(M_{a|x}\otimes\1 \, \rho_{AB}) ,\ \forall\, a,x \\
	                 					&\hspace{0.2cm} M_{a|x}\geq 0, \ \forall\, a,x \\
								& \hspace{0.2cm}  \sum_a M_{a|x} = \1, \ \forall\,x. 
	\end{split}
	\end{align}
	\end{subequations}
This SDP takes the coefficients $\{F_{a|x}\}$ of the steering inequality that were outputted by the first SDP, (\ref{seesaw1}), as input and, for the same quantum state $\rho_{AB}$, finds the set of POVMs $\{M_{a|x}\}$ that generates the assemblage that maximally violates this inequality. The measurement set that is the output of this SDP, (\ref{seesaw2}), will be the input of the first SDP, (\ref{seesaw1}), in the next round of the iteration. When performed locally on the fixed quantum state, it will necessarily generate an assemblage that has the same or a lower critical visibility than the measurement set from the previous round. When some convergence condition is satisfied (e.g., the diference between the solutions of SDP (\ref{seesaw1}), in two subsequent rounds is less than a certain value) the iteration is halted. The final value for $\eta$ found by the seesaw is an upper bound for $\eta^*(\rho_{AB},N,k)$ of the input state and the set of measurements found by the seesaw is a candidate for the optimal set of $N$ general measurements with $k$ outcomes for steering the state $\rho_{AB}$.

This is also a heuristic method, hence, by itself, it does not prove that the obtained bound is tight. However, it is possible to improve the result by testing multiple different initial points. Contrary to the search algorithm, which allows for constraints on the structure of the POVMs, this method optimizes over all possible sets of $N$ $k$-outcome general POVMs. Our calculations have shown that even though the seesaw algorithm does not have this extra feature, when the interest is in optimizing over general POVMs, it is more effective in doing so than the search algorithm (in the sense that the seesaw demands computational times that are orders of magnitude smaller than the search algorithm for the same number of measurements and outcomes). In all cases tested, for the same state and scenario the solutions of both methods coincide.

\subsection{Lower bounds for $\eta^*(\rho_{AB},N,k)$}\label{lower}

\textit{Outer polytope approximation.} Consider the set $\set{A}$ of all assemblages that can be generated by performing $N$ local measurements of $k$ outcomes on a fixed quantum state $\rho_{AB}$. This set is convex but not a polytope. In order to guarantee that all assemblages in $\set{A}$ admit an LHS model it is sufficient to guarantee that this holds for all of the extremal assemblages of the set. However, since there is an \textit{infinite number} of extremal assemblages in this set (each one corresponding to an extremal set of $N$ $k$-outcome measurements) it is not viable to test each and every one of them.

The method we propose to overcome this problem is based on the techniques presented in Ref. \cite{SuperPOVM}, where the authors approximate the set of quantum measurements by outer polytopes. The idea is to construct an external polytope $\Delta$ that contains $\set{A}$, such that every assemblage in $\set{A}$ can be expressed as a convex combination of the \textit{finitely many} extremal points of $\Delta$. We call the extremal points of $\Delta$ quasiassemblages (they are nonpositive operators that sum to a reduced state $\rho_B$). The way we generate these quasi-assemblages is by applying well-chosen quasi-POVMs \cite{SuperPOVM} (nonpositive operators that sum to the identity) to a fixed quantum state. One can calculate the white-noise robustness of each quasi-assemblage in $\Delta$ using SDP (\ref{wnrsdp}), and the lowest value among them will be a lower bound for $\eta^*(\rho_{AB}, N,k)$. The SDP will return LHS models for the quasi-assemblages which can be used to construct, by simple convex combination, LHS models for all assemblages in the depolarized set.

\begin{figure}[h!]
\begin{center}
	\includegraphics[width=5cm]{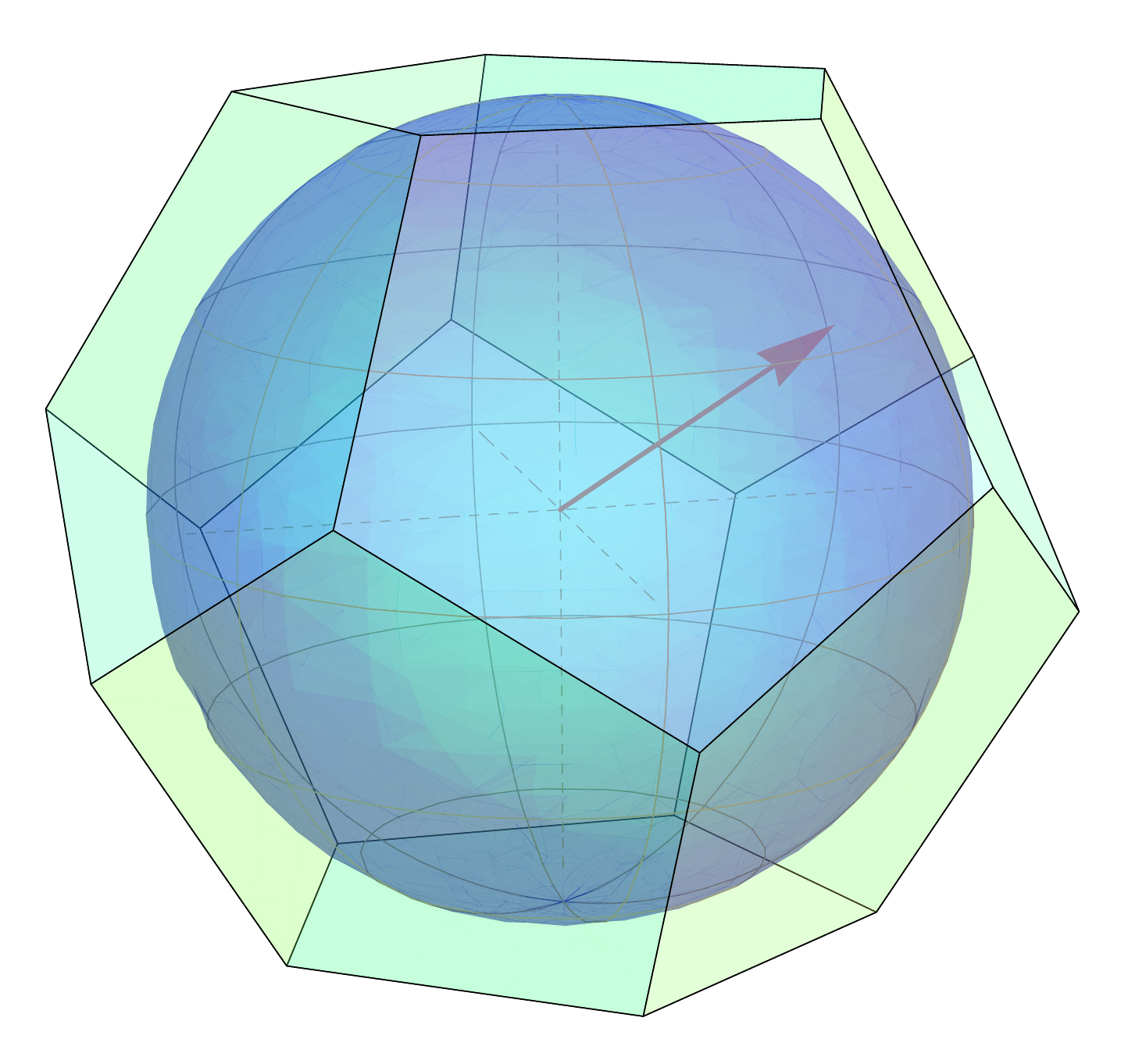}
	\caption{\small Example of an approximation of the Bloch sphere by an outer polytope.}
\label{blochpolytope}
\end{center}
\end{figure}

We now detail the construction of these polytopes for the case where the dimension of Alice's system is $d=2$. A generalization to higher dimensions follows analogously, and we refer to Ref. \cite{SuperPOVM} for more details. In this case, any measurement operator $M$ can be written as $M =\alpha\1+\vec{v}\cdot\vec{\sigma}$, where $\vec{v}$ is a three-dimensional real vector and $\vec{\sigma}$ is the vector of Pauli matrices. By checking the eigenvalues we see that $M\geq0$ if and only if $||\vec{v}||\leq\alpha$, where $||\cdot||$ is the Euclidian norm, which is equivalent to saying that $\vec{v}$ is contained in a real sphere of radius $\alpha\geq 0$. This allows one to represent each measurement operator as a vector in a rescaled Bloch sphere of radius $\alpha$.  In order to approximate the set of all POVMs in $d=2$ it is sufficient to approximate the Bloch sphere by an outer polyhedron, which is a simple task in $\mathbb{R}^3$ (see Fig. \ref{blochpolytope}).

Since the extremal points of the polytope are outside the Bloch sphere, i.e., $||\vec{v}||>\alpha$, they violate the positivity condition for operators in a two-dimensional Hilbert space. Hence, the extremal points of the polytope do not correspond to positive semidefinite operators. They are represented by a vector $v$ such that $v\cdot w_i\leq\alpha$, for some finite set $\{w_i\}$ of vectors defining facets of a polytope that contains the sphere of radius $\alpha$. Accordingly, a quasi-POVM is a set of these nonpositive operators that sum up to the identity. All sets of $N$ quasi-POVMs that can be constructed from the extremal points of the polytope that approximates the Bloch sphere are then locally performed on $\rho_{AB}$ to obtain the quasi-assemblages that define the polytope $\Delta$ that approximates the set $\set{A}$. 

The lower bound provided by this method can be improved by increasing the number of tangency points of the outer polytope on the sphere. Contrary to the search and seesaw algorithms, the outer polytope method converges to the exact value of $\eta^*(\rho_{AB},N,k)$ with probability $p=1$ in the limit of an infinite number of generic extremal points. Hence, the bound can be improved as much as necessary, up to available computational resources.

\subsection{Brief discussion of the methods}\label{methoddiscussion}

\textit{Different quantifiers of steering and joint measurability.} We start our discussion by remarking that although the presented methods were based on the white-noise robustness of steering, they can be easily adapted to estimate other quantifiers of steering that can be calculated by an SDP for fixed state and measurements. Some examples are the steering weight \cite{SteeringWeight} and the generalized robustness of steering \cite{Piani}. Also, given the strong connection between joint measurability and steering discussed in Sec. \ref{JM=STE}, analogues of all these steering quantifiers also exist for joint measurability \cite{QuantifiersDani} and our methods can be used to obtain upper and lower bounds for these quantities as well.

\textit{Convergence.} As discussed in Sec. \ref{lower}, the method to calculate lower bounds for $\eta^*(\rho_{AB},N,k)$ is constituted by a sequence of algorithms that converges to the precise value in the limit of infinite extremal points. The cumbersome feature is that the precise value cannot be attained within a finite number of steps. On the other hand, the upper-bound methods consist of heuristic optimization algorithms that may return the exact critical visibility, but there is no guarantee of that. We note that although we did not present a sequence of converging algorithms for calculating upper bounds for $\eta^*(\rho_{AB},N,k)$, one can be constructed by simply testing every possible combination of measurements, possibly with the assistance of polytopes that approximate the set of assemblages from the inside. Since the set of measurements is convex, it can be approximated by a converging sequence of polytopes, guaranteeing the existence of this sequence of algorithms  \cite{siberian}. The drawback of this  ``brute force'' converging method is that it may take an impractical amount of time to find useful bounds, which is not the case for the heuristic upper-bounds methods discussed in Sec. \ref{upper}.

\textit{Lower bounds for a finite vs. an infinite number of measurements.} In Refs. \cite{LHScavalcanti,LHShirsch}, the authors have presented a method for constructing LHS models for quantum states when all possible measurements (hence an infinite number) are considered. Here, we address a similar question, but in cases where a finite number of measurements is considered. Perhaps surprisingly, our algorithm suggests that constructing local hidden state models for only a finite number of measurements is considerably harder than for an infinite number of measurements. For instance, calculating (good) lower bounds for the critical visibility of two-qubit Werner states subjected to five dichotomic measurements was a very computationally challenging task. Nonetheless, when all possible measurements are considered, the numerical methods of Refs. \cite{LHScavalcanti,LHShirsch} can find good lower bounds in a reasonably small time.

\textit{Numerical stability of the seesaw method.} When implementing the seesaw method with the visibility parametrization described in Sec. \ref{Quantify}, we faced some numerical instability. To overcome this problem, the parametrization 
\beq
\frac{\sigma_{a|x} + t \tr(\sigma_{a|x})\frac{1}{d}}{1+t} \in LHS
\eeq 
was used instead. The SDPs were then rewritten as a minimization over the parameter $t$ with the correspondence $\eta^*=\frac{1}{1+t^*}$, where the superscript $^*$ denotes the optimal value. Although the interpretation of the visibility parametrization is more straightforward due to its relation with the depolarazing map, the formulation of the problem with the $t$ parameter is equivalent. The numerical stability was also improved by avoiding redundant constraints on normalization and nonsignaling conditions.
We also used the seesaw method to calculate upper bounds for the generalized robustness of steering \cite{Piani} of a quantum state. The seesaw for this quantifier was shown to be more numerically stable than the one for white-noise robustness. As a consequence, the generalized robustness seesaw was used to approach the scenarios with the largest number of parameters in this work.

\section{Results}

We now present the results we obtained by applying the machinery developed in the last section to some specific quantum states. In order to tackle the steering and the joint measurability problem simultaneously, we concentrate on isotropic states in our examples. Also, two-qubit isotropic states can be mapped into two-qubit Werner states \cite{WernerModel} via a local unitary transformation, which always preserves the steerability \cite{MTGeneralMeas}. For this reason, we present our two-qubit results in terms of Werner states, which in this case are given by
\beq
(\1\otimes\Lambda_\eta)(\ketbra{\psi^-}{\psi^-})=\eta\ketbra{\psi^-}{\psi^-}+(1-\eta)\frac{\1}{4},
\eeq
where $\ket{\psi^-}=\frac{1}{\sqrt{2}}(\ket{01}-\ket{10})$ is the singlet state.

To simplify notation we refer to the critical visibility of these two-qubit states as simply $\eta^*(N,k)\coloneqq\eta^*(2,N,k)$.

\subsection{Planar qubit projective measurements}\label{planar}

We start with a simple family of qubit measurements, the planar projective measurements. These are qubit projective measurements whose Bloch vectors are confined to the same plane. The reasons for studying this kind of measurement include its simple experimental implementation \cite{WisemanPlanar} and the low computational cost required to optimize over these measurements, compared to more general ones.

\begin{figure}[h!]
\begin{center}
	\includegraphics[width=\columnwidth]{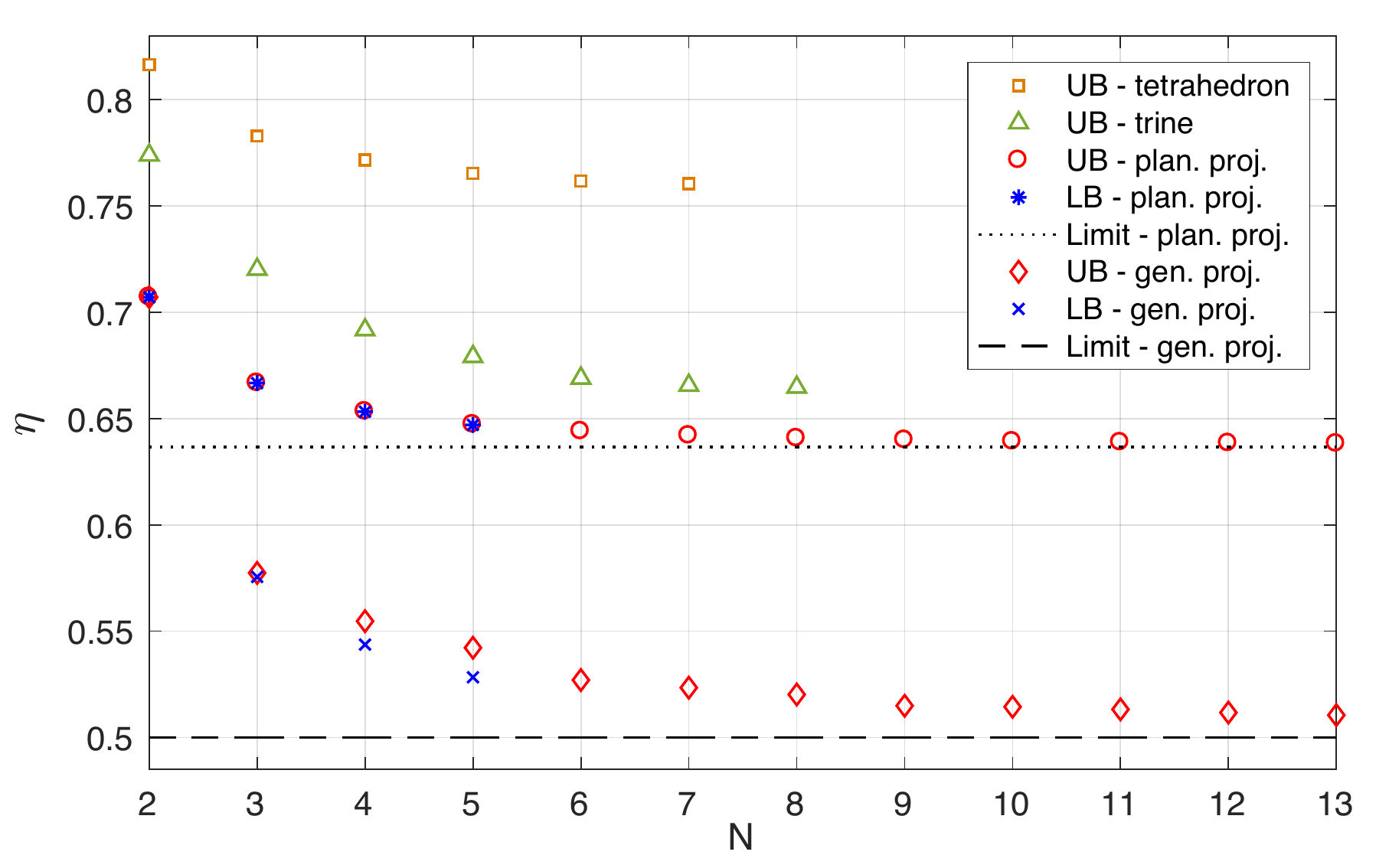}
	\caption{\small Plot of the upper bounds of the critical visibility of two-qubit Werner states subjected to $N$ regular tetrahedron and regular trine measurements and the upper and lower bounds for $N$ planar projective and general projective measurements. Dotted black lines correspond to the values $\eta=\frac{1}{2}$ bellow which two-qubit Werner states are unsteerable for all projective measurements \cite{WernerModel}, and $\eta=\frac{2}{\pi}$, bellow which two-qubit Werner states are unsteerable for all planar projective measurements \cite{WisemanPlanar,Siegen}.}
\label{graphprojective}
\end{center}
\end{figure}

Initially, we use the search algorithm with the constraint that all measurement vectors are coplanar to calculate upper bounds for the critical visibility $\eta^*(N,2)$ of two-qubit Werner states. Calculations were performed for sets of $N\in\{2,...,15\}$ planar projective measurements. The results are presented in Fig. \ref{graphprojective} and Table \ref{tablenum}.

\begin{figure}[h!]
\begin{center}
	\includegraphics[width=\columnwidth]{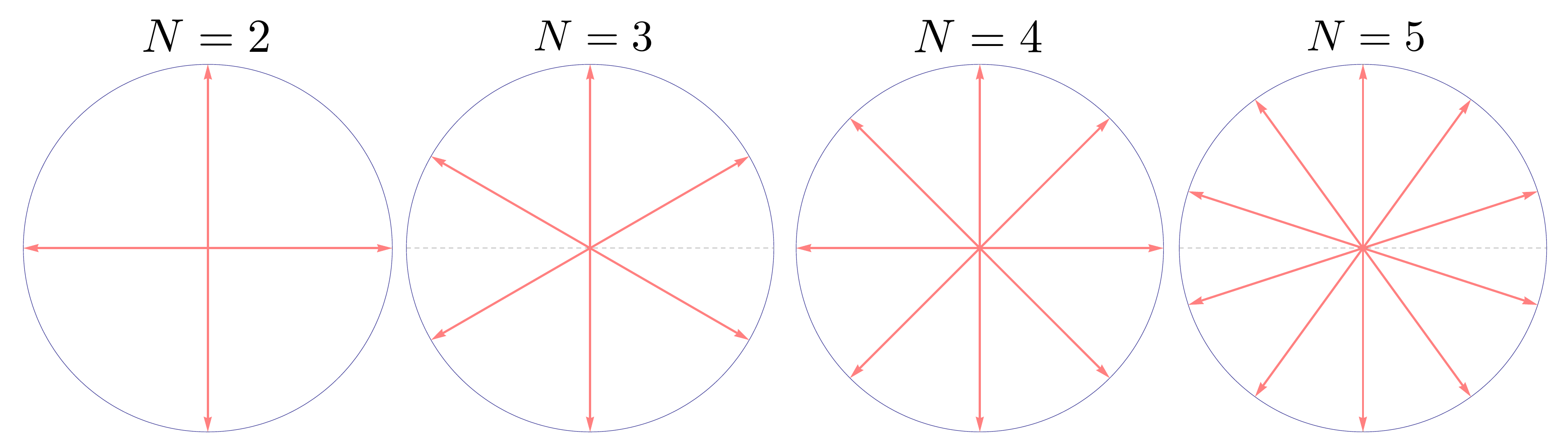}
	\caption{\small Optimal set of $N\in\{2,\ldots,5\}$ planar qubit projective measurements for steering two-qubit Werner states. Due to the connection between steering and joint measurability, these are also the most incompatible sets of $N\in\{2,\ldots,5\}$ planar qubit projective measurements.}
\label{figoptsetplan}
\end{center}
\end{figure}

For all trials performed with multiple different initial points, the result for both the objective function--the parameter $\eta$--and the optimization variables--the angles between the Bloch vectors of the measurements--were the same for all values of $N$ tested. In all cases, the optimal set of measurements found by the algorithm is the one in which the Bloch vectors of all measurements are equally spaced on a plane, i.e., each Bloch vector is separated from its next neighbors by an angle of $\frac{\pi}{N}$, as represented for the cases of $N\in\{2,...,5\}$ in Fig. \ref{figoptsetplan}.

Next, we calculated lower bounds for $\eta^*(N,2)$ in the restricted scenario of planar projective measurements using the method of outer polytope approximation. Results are reported for the cases of $N\in\{2,\ldots,5\}$ planar projective measurements also in Fig. \ref{graphprojective} and Table \ref{tablenum}. The lower bound for $\eta^*(N,2)$ found by the outer polytope approximation method matches the upper bound found by the search algorithm up to three or four decimal places for all cases tested. We consider this to be enough evidence to claim that for the cases of $N\in\{2,\ldots,5\}$ planar projective measurements, the optimal set of measurements for steering two-qubit Werner states is the set of equally spaced measurements. This is equivalent to stating that the most incompatible set of $N\in\{2,\ldots,5\}$ planar projective qubit measurements is the set of equally spaced measurements. We also conjecture this result to be valid for any number of planar projective measurements. The values we calculated match the analytical results for the incompatibility of equally spaced planar projective qubit measurements presented in Refs. \cite{WisemanPlanar,Siegen}.

\subsection{General qubit projective measurements}\label{spherical}

Since the optimal sets of measurements for our problem in the case of planar projective measurements appear to be the sets of equally spaced measurements on a plane, we hypothesize that the optimal sets of general qubit projective measurements correspond to some notion of equally spaced points on a sphere. Unfortunately, contrarily to the equivalent problem on a circumference,  the problem of equally distributing points on a sphere is not trivial and many different sets of points can be defined using different notions of distance.  This problem is particularly difficult in the regime of few points. For this work we chose the equally spaced notion of the Thomson problem \cite{Thomson} and the Fibonacci problem \cite{Fibonacci}. The former, for the particular cases of $N\in\{2,3,4,6,10\}$ projective measurements corresponding to $4,6,8,12,$ and $20$ vertices, is equivalent to the Platonic solids. The results for the critical visibility of two-qubit Werner states subjected to sets of $N\in\{2,\ldots,18\}$ local measurements constructed from these two notions of equal spacing are listed in Table \ref{tablenum}. 

\begin{table}[h!]
\begin{center}
{\renewcommand{\arraystretch}{0.75}
\begin{tabular}{| c | c c c c | c c |}               
        	       \multicolumn{7}{c}{Projective qubit measurements} \\ 
	       \hline 
	        \multicolumn{1}{|c}{} & \multicolumn{2}{c }{Gen. opt.} & \multicolumn{2}{c |}{Planar opt.} & \multicolumn{2}{c |}{Fixed sets} \\ 
                \hline \hline                 
		$N$ & Upper & Lower & Upper & Lower & Thomson & Fibonnaci \\              
		\hline 
		$\quad$ $2$ $\quad$ & $0.7071$ & $0.7071$ & $0.7071$ & $0.7071$ & $0.7071$ & $0.7102$\\			
		$3$ & $0.5774$ & $0.5755$ & $0.6667$ & $0.6667$ & $0.5774$ & $0.6981$ \\	
		$4$ & $0.5547$ & $0.5437$ & $0.6533$ & $0.6532$ & $0.5774$ & $0.6114$  \\	
		$5$ & $0.5422$ & $0.5283$ & $0.6472$ & $0.6470$ & $0.5513$ & $0.5653$  \\
		$6$ & $0.5270$ & $$ & $0.6440$ & $$ & $0.5393$ & $0.5561$  \\
		$7$ & $0.5234$ & $$ & $0.6420$ & $$ & $0.5234$ & $0.5533$ \\
		$8$ & $0.5202$ & $$ & $0.6407$ & $$ & $0.5250$ & $0.5508$ \\
		$9$ & $0.5149$ & $$ & $0.6399$ & $$ & $0.5209$ & $0.5359$  \\
		$10$ & $0.5144$ & $$ & $0.6392$ & $$ & $0.5191$ & $ 0.5302$  \\
		$11$ & $0.5132$ & $$ & $0.6388$ & $$ & $0.5148$ & $0.5274$  \\
		$12$ & $0.5117$ & $$ & $0.6384$ & $$ & $0.5152$ & $0.5261$  \\
		$13$ & $0.5105$ & $$ & $0.6382$ & $$ & $0.5126$ & $0.5220$  \\
		$14$ & $$ & $$ & $ 0.6380$ & $$ & $0.5114$ & $ 0.5180$  \\
		$15$ & $$ & $$ & $0.6378$ & $$ & $ 0.5107$ & $0.5158$  \\	
		$16$ & $$ & $$ & $$ & $$ & $0.5106$ & $0.5158$  \\
		$17$ & $$ & $$ & $$ & $$ & $0.5086$ & $0.5150$  \\
		$18$ & $$ & $$ & $$ & $$ & $0.5079$ & $0.5136$  \\		
           	\hline 
\end{tabular}
}
\end{center}
\caption{Summary of numerical results for the critical visibility of two-qubit Werner states subjected to $N$ projective measurements.}
\label{tablenum}
\end{table}

\begin{figure}[h!]
\begin{center}
	\includegraphics[width=\columnwidth]{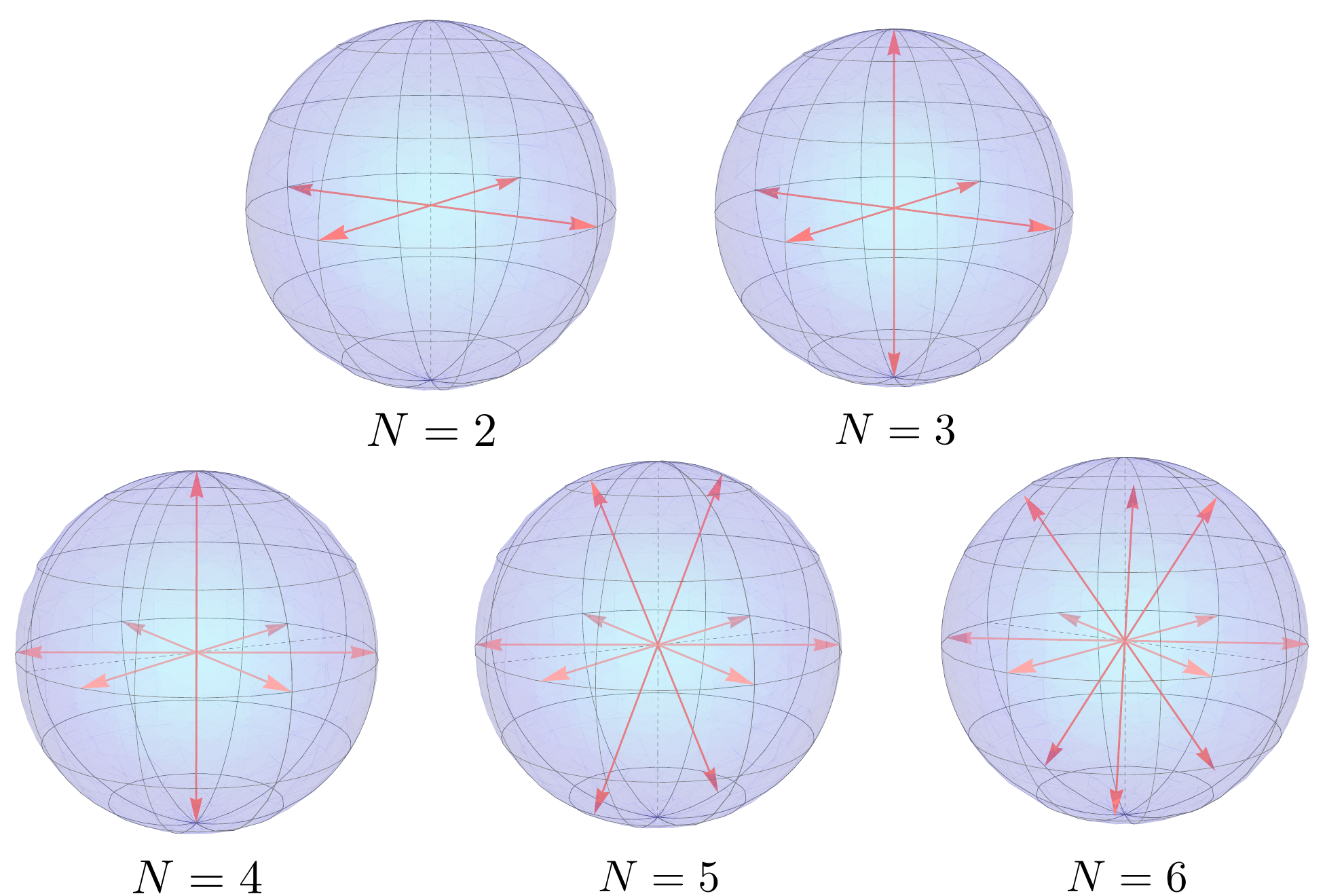}
	\caption{\small Candidates for the optimal set of $N\in\{2,\ldots,6\}$ qubit measurements for steering two-qubit Werner states. These sets are also candidates for the most incompatible set of $N\in\{2,\ldots,6\}$ qubit measurements.}
\label{figoptsetsph}
\end{center}
\end{figure}

To test whether these sets of measurements are indeed optimal, we once again use the search algorithm, now with the only restriction that the measurement operators correspond to projectors. We report upper bounds for the value of $\eta^*(N,2)$ in scenarios of $N\in\{2,\ldots,13\}$ general projective measurements. In all cases the search algorithm was able to improve the bound provided by both the Thomson and the Fibonacci measurements (see Table \ref{tablenum}), proving that they are actually not optimal. The best upper bounds are plotted in Fig. \ref{graphprojective} and the Bloch vectors of the measurement elements that form the best candidate for the optimal set of measurements in the cases of $N\in\{2,\ldots,6\}$ projective measurements are plotted in Fig. \ref{figoptsetsph}.

For these measurements, the vectors are distributed in a particular way: for $2$ and $3$ measurements, we have sets of orthogonal vectors; for $4$ measurements we have $3$ coplanar and equally distributed vectors and $1$ vector orthogonal to the other $3$; for $5$ and $6$ measurements, the structure of $3$ coplanar equally spaced vectors is maintained and the other vectors are agglomerated around the poles of the sphere with the same $z$-projection. For the cases of $7$ or more measurements, this apparent symmetry is no longer necessarily respected.

Using the outer polytope approximation we calculated the lower bounds for the cases of $N\in\{2,\ldots,5\}$ projective measurements that can be seen on Fig. \ref{graphprojective} and Table \ref{tablenum}. In this case the gap between upper and lower bounds for the general projective case is larger than for the planar projective case. This is due to the increase in the number of parameters in the former case as compared to the latter. However, due to the convergence properties of our outer polytope method, as discussed on Secs. \ref{lower} and \ref{methoddiscussion}, these lower bounds can be improved beyond what is the scope of this work.

\subsection{General POVM relevance for qubits}\label{povm}

An old standing question in nonlocality is to understand when general POVMs are useful to reveal this property in a given quantum state \cite{WernerModel,Barrett,WernerPOVM,nguyen16}. It is well known that constructing local hidden variable (LHV) and LHS models for general POVMs is considerably harder than constructing these models for projective measurements \cite{WernerModel,Barrett,WernerPOVM,nguyen16}. Moreover, it is not known whether nonprojective measurements are more useful than projective ones to demonstrate EPR steering or Bell nonlocality. For some particular fixed (nontight) Bell inequalities, it is known that general POVMs can lead to a larger Bell violation than projective measurements \cite{RelevantPOVM}, but the existence of a quantum state that has a LHV/LHS model for projective measurements but displays Bell nonlocality/EPR steering when general POVMs are considered is still an open question. 

We have applied our seesaw method to two-qubit Werner states where the noncharacterized party has access to $N\in\{2,\ldots,7\}$ general POVMs with $2,3,$ and $4$ outcomes. We recall that for the case of $2$ outputs, nonprojective POVMs can never be useful for nonlocality, since they can always be written as convex combinations of projective measurements \cite{ExtremalPOVM}. Also, qubit POVMs with more than $4$ outcomes are never extremal \cite{ExtremalPOVM}, hence these measurements could never lead to better bounds for the critical visibility. For this reason we now define the quantity $\eta^*(N)\coloneqq\eta^*(N,d^2=4)$, the critical visibility of two-qubit Werner states when subjected to $N$ POVMs of an arbitrary number of outcomes. 

\begin{figure}[h!]
\begin{center}
	\includegraphics[width=7cm]{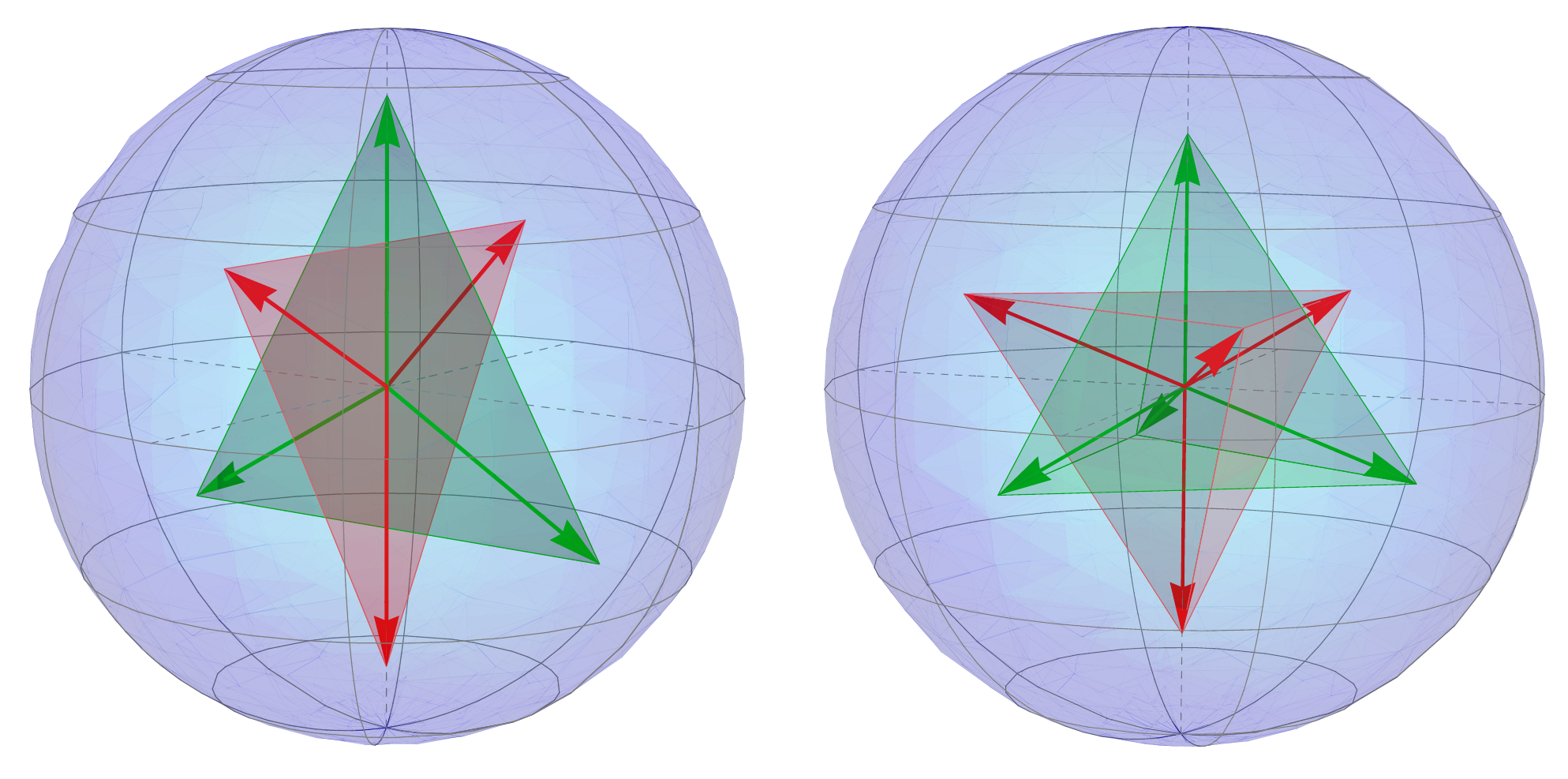}
	\caption{\small Candidates for the optimal set of $N=2$ regular trine (left; $k=3$) and regular tetrahedron (right; $k=4$) symmetric qubit POVMs for steering two-qubit Werner states.}
\label{optpovm}
\end{center}
\end{figure}

In addition to using the seesaw method to explore general POVMs, we have applied the search algorithm to the specific case where Alice is required to perform symmetric $3$- and $4$-outcome POVMs on her side of a maximally entangled two-qubit state. In the $4$-outcome case, we have fixed all measurements to be SIC-POVMs \cite{renes04}, which are extremal measurements whose Bloch vectors correspond to the vertices of a regular tetrahedron. In the $3$ outcomes case, the chosen POVM was the symmetric extremal measurement whose Bloch vectors correspond to the ``Mercedes-Benz star'', also called the regular trine \cite{Trine} (see Fig. \ref{optpovm}). These particular symmetric nonprojective measurements are known to be useful in tasks such as tomography \cite{rehacek04,SICPOVMTomography} and cryptography \cite{fuchs03}, hence they are interesting examples of extremal nonprojective qubit POVMs \cite{ExtremalPOVM}.

\begin{table}[h!]
\begin{center}
{\renewcommand{\arraystretch}{0.75}
\begin{tabular}{| c | c c  c |}
		\multicolumn{4}{c}{Symmetric qubit POVMs} \\ 
		\hline                             
		$N$ & Proj. ($k=2$) & Trine ($k=3$) & Tetra. ($k=4$) \\              
		\hline 
		$\quad$ $2$ $\quad$ & $\quad$ $0.7071$ $\quad$ & $\quad$ $0.7739$ $\quad$ &  $\quad$ $0.8165$ $\quad$\\	
		$3$ & $0.5774$ & $0.7202$  & $0.7829$ \\	
		$4$ & $0.5547$ & $0.6917$  & $0.7716$ \\
		$5$ & $0.5422$ & $0.6791$  & $0.7653$ \\
		$6$ & $0.5270$ & $0.6690$  & $0.7617$ \\
		$7$ & $0.5234$ & $0.6656$  & $0.7605$ \\
		$8$ & $0.5202$ & $0.6647$ & -- \\
              	\hline
\end{tabular}
}
\end{center}
\caption{Summary of numerical results for upper bounds of the critical visibility of two-qubit Werner states subjected to $N$ extremal symmetric POVMs.}
\label{tablesic}
\end{table}

Our results for symmetric qubit POVMs are plotted in Fig. \ref{graphprojective} and listed in Table \ref{tablesic} for the cases of $N\in\{2,\ldots,8\}$, including the results for projective measurements, which are symmetric $2$-outcome POVMs, for the sake of comparison. In the case of $N=2$, the optimal set of regular trine and regular tetrahedron POVMs is plotted in Fig. \ref{optpovm}. It is easy to see that under none of the analyzed scenarios were the symmetric nonprojective POVMs able to show more steering than the projective measurements. In fact, the bounds for symmetric $3$- and $4$-outcome qubit POVMs are considerably worse than for projective qubit measurements.

As for the optimization over general $3$- and $4$-outcome qubit POVMs, we could \textit{not} find any set of $N$ general POVMs that are able to overperform projective ones. For $N=2$ and $3$ and $k=4$, the seesaw algorithm ran $10^5$ times, each time with a different initial point; for $N=4$ and $k=3$, the seesaw ran $4\times10^4$ times, and for $k=4$, $3\times10^4$ times; for $N=5$, $6$, and $7$, and $k=3$, it ran $2\times10^4$, $2\times10^3$, and $200$ times, respectively.  Without exception, our algorithm recovered the bound for $\eta^*(N)$ obtained by the optimization over projective measurements using general POVMs, usually by nulling two measurement outcomes and ``simulating'' a projective measurement. However, it was never able to surpass it.

Strictly speaking, the results presented in this section are only upper bounds for the critical visibility $\eta^*(N)$. Nevertheless, given the small number of parameters in the two-qubit scenario and the number of times we have ran our heuristic method, we believe that these results are strong evidence that general POVMs are not useful to reveal EPR-steering in two-qubit Werner states.

\subsection{Higher dimension states and measurements}\label{higherd}

We now explore the generality of our seesaw method on quantum systems of dimension $d>2$ by calculating bounds for the critical visibility  $\eta^*(d,N,k)$ of higher dimension maximally entangled states. Let us start with the simple case where these states are subjected to only $2$ local general measurements. These calculations are reported for states of dimensions $d\in\{2,\ldots,6\}$ in Table \ref{tablen2}. We note that by increasing the number of outcomes in the measurements from $k=2$ up to $k=d$ the bounds for $\eta^*(d,2,k)$ are significantly improved and the optimal sets of measurements are always composed by projective measurements--even though most of these scenarios allow extremal nonprojective POVMs. However, once the number of outcomes achieves $k=d+1$ the bound for $\eta^*(d,2,k)$ provided by the seesaw method ceases to decrease and it seems that increasing the number of outputs beyond this point does not improve the results. Since there only exist projective measurements with up to $k=d$ outcomes, this result is evidence that allowing POVMs more general than projective measurements does not increase the robustness of the steerability of isotropic states. Following the connection between the steerability of these states and joint measurability, this is also evidence that sets of $2$ general qudit POVMs cannot be more incompatible than sets of $2$ projective qudit measurements.

\begin{table}[h!]
\begin{center}
{\renewcommand{\arraystretch}{0.75}
\begin{tabular}{| c | c c c c c |}  
        		\multicolumn{6}{c}{$N=2$} \\
        		\hline              
		$\quad$ $k$ $\quad$& $d=2$ & $3$ & $4$ & $5$ & $6$  \\		
		\hline	
		$2$ & $0.7071$ & $0.7000$ & $0.6901$ & $0.6812$ & $0.6736$ \\	
		$3$ & $0.7071$ & $0.6794$ & $0.6722$ & $0.6621$ & $0.6527$ \\	
		$4$ & $$ & $0.6794$ & $0.6665$ & $0.6544$ & $0.6448$ \\
		$5$ & $$ & $$ & $0.6665$ & $0.6483$ & $0.6429$ \\	
		$6$ & $$ & $$ & $$ & $0.6483$ & $0.6390$  \\	
		$7$ & $$ & $$ & $$ & $$ & 0.6390 \\		
              	\hline
\end{tabular}
}
\end{center}
\caption{Summary of numerical results for upper bounds of the critical visibility of $d$-dimensional isotropic states subjected to $2$ general POVMs of $k\in\{2,\ldots,d+1\}$ outcomes.}
\label{tablen2}
\end{table}

Since the scenario where the uncharacterized party is allowed to perform only $2$ measurements is very particular, we performed the same calculations reported above for $d$-dimensional isotropic states allowing scenarios with $3$ and $4$ general measurements with outcomes up to $d+1$ as well. In these broader scenarios, we also calculated upper bounds for the critical visibility $\eta^*(d,N,k)$ of the isotropic states. However, since the number of parameters increases too rapidly (exponentially on the number of measurements), the seesaw method presented some numerical instability, and for this reason we are not able to reach any conclusions about the relevance of general POVMs in these scenarios.

\subsection{Mutually unbiased bases}\label{MUB}

A set of MUBs consists of 2 or more orthonormal bases $\{\ket{a_x}\}_a$ in a $d$-dimensional Hilbert space that satisfy
\begin{equation}
|\braket{a_x}{b_y}|^2=\frac{1}{d}, \quad \forall\, a,b \in\{1,\ldots,d\}, \ x\neq y,
\end{equation}
for all bases $x,y$ \cite{durt10}. A set of MUBs is called complete if for a Hilbert space of dimension $d$ there exists $d+1$ MUBs. These bases can be used to construct sets of mutually unbiased projective measurements with a high level of symmetry, and for this reason one might think they would be good candidates for the optimal set of measurements for measurement incompatibility and for EPR steering with a maximally entangled state. 

We have calculated the critical visibility of isotropic states of dimension $d\in\{2,\ldots,6\}$ when subjected to local MUB measurements using SDP (\ref{wnrsdp}) and listed the results in Table \ref{tablemub}. These exact values calculated by our SDP (\ref{wnrsdp}), show significant improvement over the analytical bounds obtained in Refs. \cite{marciniak15,hgsieh16} for steering with MUBs and maximally entangled states. Next, we used the seesaw method to calculate upper bounds for $\eta^*(d,N,d)$ of the isotropic states when locally subjected to sets of general POVMs with $d$ outcomes for some number of measurements $N$ where MUB measurements are known to exist. Perhaps surprisingly, in many cases we found sets of measurements with greater or equal robustness, showing that MUBs are not necessarily the best choice of measurements to reveal quantum steering nor are they the most incompatible ones. The results are listed in Table \ref{tablemub}. The optimal measurements found by the seesaw method are all projective measurements in these cases as well. We remark that in Refs. \cite{carmeli11,heinosaari13}, the authors have computed (analytically) the required visibility $\eta^*(d,2,d)$ for any pair of $d$-dimensional MUB measurements to be jointly performed; here we have shown that there exist pairs of measurements that are even more incompatible than mutually unbiased ones. However, in scenarios where there exist complete sets of MUB measurements, for dimensions $2$, $3$, and $4$, we were not able to find measurements more resistant to white noise and better for steering isotropic states than the MUB ones, which is evidence that they may be optimal for this task.

\begin{table}[h!]
\begin{center}
{\renewcommand{\arraystretch}{0.75}
\begin{tabular}{| c | c c c c c |}  
        		\multicolumn{6}{c}{MUBs} \\
        		\hline              
		$\quad$ $N$ $\quad$ & $d=2$ & $3$ & $4$ & $5$ & $6$  \\		
		\hline	
		$2$ & $0.7071$ & $0.6830$ & $0.6667$ & $0.6545$ & $0.6449$ \\	
		$3$ & $0.5774$ & $0.5686$ & $0.5469$ & $0.5393$ & $0.5204$ \\	
		$4$ & $$ & $0.4818$ & $0.5000$ & $0.4615$ & $$ \\
		$5$ & $$ & $$ & $0.4309$ & $0.4179$ & $$ \\	
		$6$ & $$ & $$ & $$ & $0.3863$ & $$  \\			
              	\hline
		\multicolumn{6}{c}{} \\
	         \multicolumn{6}{c}{General $d$-outcome POVMs} \\
        		\hline              
		$N$& $d=2$ & $3$ & $4$ & $5$ & $6$  \\		
		\hline	
		$2$ & $0.7071$ & $0.6794$ & $0.6665$ & $0.6483$ & $0.6395$ \\	
		$3$ & $0.5774$ & $0.5572$ & $0.5412$ & $0.5266$ & $0.5139$ \\	
		$4$ & $$ & $0.4818$ & $0.4797$ & $0.4615$ & $$ \\
		$5$ & $$ & $$ & $0.4309$ & -- & $$ \\	
		$6$ & $$ & $$ & $$ & -- & $$  \\	
		\hline
\end{tabular}
}
\end{center}
\caption{Comparison between the exact critical visibility of isotropic states in dimension $d$ subject to local mutually unbiased measurements and the upper bound of the same states when optimizing over general POVMs with $k=d$.}
\label{tablemub}
\end{table}

\section{Discussion}

We have used three methods for investigating EPR steering and joint measurability under restrictive measurement scenarios and discussed the applicability of each one. Using white-noise robustness as a quantifier, we have presented two heuristic methods for calculating the critical visibility of quantum states subjected to a finite number of measurements and one converging method for lower-bounding the same quantity. Our methods can be easily adapted to other steering and joint measurability quantifiers.

For two-qubit Werner states, we showed that the best sets of $N\in\{2,\ldots,5\}$ planar projective measurements are equally spaced measurements and conjecture this result to be valid for all $N\in\mathbb{N}$. Our upper bounds for the critical visibility of two-qubit Werner states subjected to planar projective measurements match the analytical expressions derived in Refs. \cite{WisemanPlanar,Siegen} for equally spaced measurements. We proved that intuitive notions of equally spaced measurements in the Bloch sphere, like the vertices of Platonic solids, do not correspond to the best measurements to show steering with two-qubit Werner states, nor are they the most incompatible sets of measurements. We showed that symmetric $3$- and $4$-outcome qubit POVMs are not optimal for steering two-qubit Werner states as well. Upper bounds for the critical visibility of two-qubit Werner states subjected to $N\in\{2,\ldots,18\}$ general measurements were calculated. We provided strong numerical evidence that general POVMs are not more suitable for steering two-qubit Werner states than projective measurements, and suggested candidates for the optimal sets of $N\in\{2,\ldots,6\}$ qubit measurements that are projective and follow a nonintuitive pattern.

Our results for higher dimension isotropic states indicate that increasing the number of outcomes until $k=d$ improves the bound for the critical visibility of the state. However, increasing the number of outcomes beyond the value of the local dimension of the state does not seem to improve the bounds, which strengthens the idea that nonprojective POVMs are not relevant for steering. The candidates for optimal measurements in all higher dimension scenarios are projective measurements. Finally, we proved that many incomplete sets of MUB measurements are not optimal for steering and provided numerical evidence that complete sets of MUB measurements could be optimal for steering isotropic states.

Although we presented numerical evidence against the relevance of nonprojective POVMs for EPR steering, deciding if projective measurements are indeed optimal for steering in all scenarios and for all quantum states still remains an open question. 
One future direction is to apply similar techniques for the study of Bell nonlocality. Although some simple adaptation of our methods can be used to tackle the analogous problem for Bell nonlocality, the number of parameters in the problem could make our algorithms impracticable even in simple scenarios.

All code written for this work is available in a repository \cite{Code}.

\begin{acknowledgements}
The authors thank Teiko Heinosaari, Peter Wittek, and Paul Skrzypczyk for interesting discussions. This work was supported by the Brazilian agencies CAPES, CNPq, and FAEPEX, the Austrian Science Fund (FWF) through the START project Y879-N27, the Japan Society for the Promotion of Science (JSPS) through KAKENHI Grant No. 16F16769, a Ramon y Cajal fellowship, the Spanish MINECO (QIBEQI FIS2016-80773-P and Severo Ochoa SEV-2015-0522), Fundació Cellex, Generalitat de Catalunya (SGR875 and CERCA Program), and ERC CoG QITBOX.
\end{acknowledgements}

\bibliographystyle{jess_unsrt}
{\small{\bibliography{biblio.bib}}}

\end{document}